\documentclass{article}
\usepackage{amssymb}
\usepackage{amsmath}
\def\R{{\mathbb R}}

\usepackage{graphicx}

\begin{document}

\title{The minimal model of Hahn for the Calvin cycle}

\author{Hussein Obeid and Alan D. Rendall\\
Institut f\"ur Mathematik\\
Johannes Gutenberg-Universit\"at\\
Staudingerweg 9\\
D-55099 Mainz\\
Germany}

\date{}

\maketitle 

\begin{abstract}
There are many models of the Calvin cycle of photosynthesis in the literature.
When investigating the dynamics of these models one strategy is to look at the 
simplest possible models in order to get the most detailed insights. We
investigate a minimal model of the Calvin cycle introduced by Hahn while he
was pursuing this strategy. In a variant of the model not including 
photorespiration it is shown that there exists exactly one positive steady 
state and that this steady state is unstable. For generic initial data
either all concentrations tend to infinity at lates times or all
concentrations tend to zero at late times. In a variant including 
photorespiration it is shown that for suitable values of the parameters of the
model there exist two positive steady states, one stable and one unstable.
For generic initial data either the solution tends to the stable steady state
at late times or all concentrations tend to zero at late times. Thus we 
obtain rigorous proofs of mathematical statements which together confirm
the intuitive idea proposed by Hahn that photorespiration can stabilize the
operation of the Calvin cycle. In the case that the concentrations tend to
infinity we derive formulae for the leading order asymptotics using the 
Poincar\'e compactification. 
\end{abstract}

\section{Introduction}

The Calvin cycle is a part of photosynthesis. There are many mathematical
models for this biochemical system in the literature. Reviews of these
can be found in \cite{arnold11}, \cite{arnold14} and \cite{jablonsky11}.
This is an interesting example in which the relations between different 
mathematical models for the same biological situation can be investigated.
A mathematical comparison of a number of these models was carried out in
\cite{rendall17}. There it was pointed out that it would be desirable to 
look more closely at the minimal model of the Calvin cycle introduced by
Hahn \cite{hahn91}. In fact Hahn's paper contains several related systems
of ordinary differential equations of dimensions two and three and the aim
of the present paper is to obtain an understanding of the dynamics of the 
two-dimensional models of Hahn which is as complete as possible. There is
also a brief discussion of the relation of the two-dimensional models to the
three-dimensional one.

The function of the Calvin cycle is to use carbon dioxide to produce sugars.
This process is fuelled by ATP and NADPH produced in the light reactions of
photosynthesis where the energy contained in light is captured as chemical
energy. A comprehensive introduction to the biochemistry of photosynthesis
can be found in \cite{heldt11}. The main step in the Calvin cycle, resulting in
the production of PGA (phosphoglycerate), is catalysed by the enzyme Rubisco.
Interestingly this enzyme has a dual functionality. It can not only catalyse
the reaction of carboxylation, which is the primary way in which carbon
dioxide is fixed in the Calvin cycle, but also an oxidation reaction. This
second reaction competes with the first and reduces the efficiency with which
the Calvin cycle produces sugar. The reason for the existence of this 
apparently wasteful alternative reaction is not clear. One 
possible explanation, for which the Hahn model is relevant, is that 
photorespiration stabilizes the system - it creates the possibility of the
existence of a stable positive steady state. 

The paper is organized as follows. In section \ref{hahn} the two-dimensional
system of Hahn is introduced. In dimensionless form the equations depend on
two non-negative parameters $\alpha$ and $\beta$. The case $\beta>0$ 
corresponds to including photorespiration in the model. The dynamics of the 
models is first analysed in the case without photorespiration ($\beta=0$). 
The main result is Theorem 1 which describes the global asymptotic behaviour
of general solutions in detail. There exists a unique positive steady state
$S_1$ which is unstable. For an open set of initial data which is described in 
detail all concentrations tend to zero as $t\to\infty$. For another open 
set of initial data all concentrations tend to infinity as $t\to\infty$.
The complement of the union of these two sets is the stable manifold of 
the steady state $S_1$. A formula is derived for the leading order asymptotics
of the solutions which tend to infinity. In section \ref{photoresp} the
case $\beta>0$ is treated. The main result is Theorem 2. In one open set of
parameter space, for which an explicit formula is given, all solutions have 
the property that the concentrations tend to zero as $t\to\infty$. In the
interior of the complement of that set more interesting behaviour is observed.
There are two positive steady states, one stable and one unstable. For an open
set of initial data all concentrations tend to zero as $t\to\infty$. For 
another open set of initial data the solutions tend to the stable positive
steady state as $t\to\infty$.

Many models of the Calvin cycle contain the fifth power of the concentration
of the substance GAP (glyceraldehyde phosphate). This is because in the usual
coarse-grained descriptions of the Calvin cycle, where many elementary reactions
are combined, there is an effective reaction where five molecules of GAP
with three carbon atoms each go in and three molecules of a five-carbon sugar
come out. Applying mass-action kinetics to this leads to the fifth power. In
deriving the model studied in sections \ref{hahn} and \ref{photoresp} Hahn
replaces the fifth power by the second power. His motivation is to make the
model analytically more tractable. He assumes implicitly that this change 
makes no essential difference to the qualitative behaviour of the solutions
but gives no justification for this assumption. In section \ref{fifth} we show
that the solutions of the model with the fifth power do indeed behave in a way
which is very similar to the behaviour of the model with the second power. The
main difference in the analysis is that for the fifth power no explicit 
formula is obtained for the boundary between the two generic behaviours in 
parameter space. The results are summarized in Theorems 3 and 4. In 
\cite{hahn91} the 
two-dimensional systems are obtained from a three-dimensional one by informal 
arguments. In section \ref{3d} it is shown how the relation between the 
three-dimensional system and the two-dimensional system with the fifth power 
can be formalized in a rigorous way using the theory of fast-slow systems. 
(For an introduction to this theory we refer to \cite{kuehn15}.) This also 
gives some limited information about the dynamics of solutions of the 
three-dimensional system. A full analysis of the three-dimensional system is a 
much harder problem which is left to future work. 

\section{The system of Hahn}\label{hahn}

The system which will be examined in what follows consists of the
equations (41)-(42) in \cite{hahn91}:
\begin{eqnarray}
&&\frac{dx}{dt}=-\alpha x-2\beta x^2+3y^2,\label{hahn1}\\
&&\frac{dy}{dt}=2\alpha x+3\beta x^2-5y^2-y.\label{hahn2}
\end{eqnarray}
In addition to the system with $\alpha>0$ and $\beta>0$, which we will call
the full system, we also treat the cases where $\alpha=0$ (no photosynthesis),
$\beta=0$ (no photorespiration) or both. Note for future reference that the
derivative of the right hand side of this system at the point $(x,y)$ is
\begin{equation}
\left[
{\begin{array}{cc}
-\alpha-4\beta x & 6y\\ 
2\alpha+6\beta x & -10y-1 \\ 
\end{array}}
\right]
\end{equation}
with determinant $\alpha+4\beta x-2\alpha y+4\beta xy$.

Consider first the case $\alpha=\beta=0$. There any solution satisfies the
inequality $\frac{dy}{dt}\le -y$ and thus $y$ decays exponentially at late 
times. In particular there are no positive steady states. The non-negative 
steady states are precisely the points on the $x$-axis. Apart from the zero 
eigenvalue due to the continuum of steady states the other eigenvalue of the 
linearization about any of these points is $-1$ and this manifold is normally 
hyperbolic \cite{kuehn15}. It follows that given any $x_0>0$ there exists a 
positive solution with $\lim_{t\to\infty}x(t)=x_0$. Consider next the case 
$\alpha=0$, $\beta\ne 0$. Any positive steady state satisfies 
$y=\sqrt{\frac{2\beta}{3}}x$ by (\ref{hahn1}). Substituting this into 
(\ref{hahn2}) gives $y(-\frac12 y-1)=0$. Thus there is no positive steady 
state. The only non-negative steady state is at the origin. In fact 
$\frac{d}{dt}(3x+2y)=-y^2-y$. Thus $3x+2y$ is a strict Lyapunov function on 
the non-negative orthant away from the origin and it follows that all 
solutions converge to the origin as $t\to\infty$.  

\begin{figure}[ht]
\begin{center}
\includegraphics[scale=0.4]{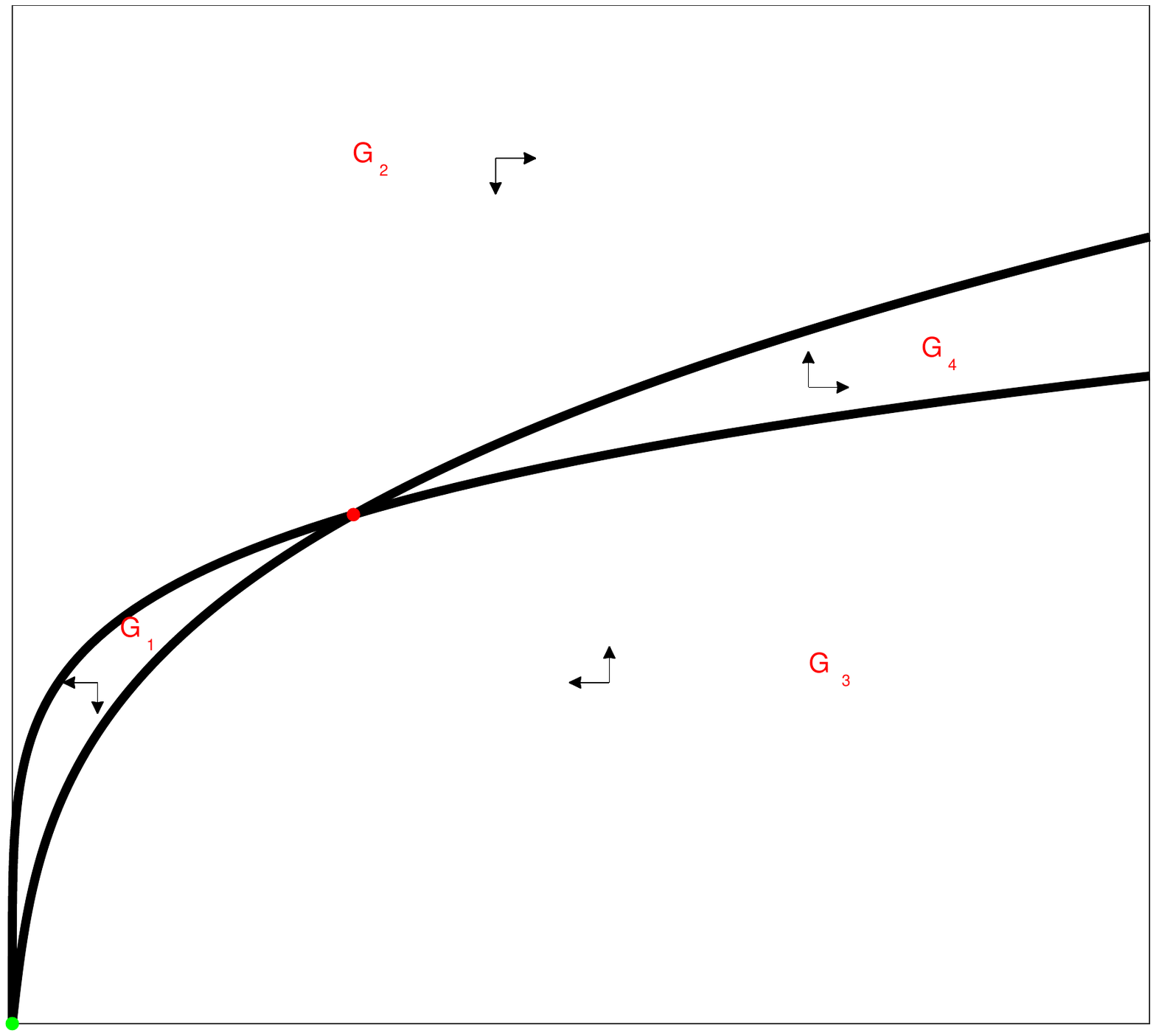}
\end{center}
\caption{Phase portrait in the absence of photorespiration}
\label{fig:nophotoresp}
\end{figure}

In the case $\alpha\ne 0$, $\beta=0$ we have the inequality 
$\frac{d}{dt}(5x+3y)\le\frac15\alpha (5x+3y)$, so that all solutions exist
globally in the future. Equation (\ref{hahn1}) shows that for a
steady state $x=\frac{3}{\alpha}y^2$. Substituting this into (\ref{hahn2}) 
gives $y^2-y=0$. Thus the steady states are $S_0=(0,0)$ and  
$S_1=\left(\frac{3}{\alpha},1\right)$. Now we carry out a nullcline 
analysis as described in the Appendix. The nullclines are given by 
$x=\frac{1}{\alpha}y^2$ and $x=\frac{1}{2\alpha}(5y^2+y)$. These are the graphs
of functions of $y$ and it is clear that the complement of the union of the 
nullclines has four connected components (cf. Figure~\ref{fig:nophotoresp}). 
These are 
\begin{eqnarray}
&&G_1=(-,-)=\{(x,y):x>\frac{3}{\alpha}y^2,x<\frac{1}{2\alpha}(5y^2+y),\\
&&G_2=(+,-)=\{(x,y):x<\frac{3}{\alpha}y^2,x<\frac{1}{2\alpha}(5y^2+y),\\
&&G_3=(-,+)=\{(x,y):x>\frac{3}{\alpha}y^2,x>\frac{1}{2\alpha}(5y^2+y),\\
&&G_4=(+,+)=\{(x,y):x<\frac{3}{\alpha}y^2,x>\frac{1}{2\alpha}(5y^2+y).
\end{eqnarray}
The complement of $S_1$ in one of the nullclines has two connected
components which can be distinguished by the sign of the time derivative
which does not vanish. We write 
\begin{eqnarray}
&&N_1=\{0,0\}\cup (0,-)\cup S_1\cup (0,+),\\
&&N_2=\{0,0\}\cup (-,0)\cup S_1\cup (+,0).
\end{eqnarray}
Note that if $\dot x=0$ at some time then $\ddot x=6y\dot y$ and that if 
$\dot y=0$ at some time then $\ddot y=(2\alpha+6\beta x)\dot x$. 

\noindent
{\bf Lemma 1} A solution of (\ref{hahn1})-(\ref{hahn2}) belongs to one 
of the following three cases.

\noindent
(i) It converges to $S_0$ as $t\to\infty$.

\noindent
(ii) It converges to $S_1$ as $t\to\infty$.

\noindent
(iii) There is a time $t_1$ such that it belongs to $G_4$ for $t\ge t_1$.

\noindent
{\bf Proof} Consider a solution which starts at a point on the boundary of 
$G_1$ other than $S_0$ or $S_1$. If it is on $N_1$ then $\dot y<0$ and 
using the equation for $\ddot x$ shows that $\dot x$ immediately becomes 
negative. If a solution starts on $N_2$ then $\dot x<0$ and $\dot y$
immediately becomes negative. It follows that any solution which starts in 
$G_1$ remains in $G_1$ and any solution which starts at a point on the 
boundary of $G_1$ other than a steady state immediately enters $G_1$. Since 
$y$ is decreasing for solutions in $G_1$ any solution which is ever in $G_1$ 
converges to the origin as $t\to\infty$. Consider next a solution which starts 
in $G_2$. Since $y$ is decreasing on $G_2$ this solution is bounded. By
Lemma A1 of the Appendix it either converges to $S_0$ or to $S_1$ as 
$t\to\infty$ or it reaches another point of the boundary of $G_2$ after a 
finite time. In the latter case it reaches a point of the boundary of $G_1$ or 
$G_4$ after a finite time. By an analogous argument we can reach a similar 
conclusion about a solution which starts in $G_3$. Either it converges to $S_0$ 
or $S_1$ or it reaches another point of the boundary of $G_1$ or
$G_4$ after a finite time. An analysis similar to that done for $G_1$
can be carried out for $G_4$. A solution which starts in $G_4$ must
remain there and a solution which starts on the boundary of $G_4$ must
immediately enter $G_4$. Thus any solution which does not belong to case (i)
or case (ii) must enter $G_4$ after finite time and then it stays there. 
$\blacksquare$   

\noindent
{\bf Lemma 2} The stable manifold of $S_1$ intersects both axes. If a solution 
starts below the stable manifold of $S_1$ it converges to $S_0$ as 
$t\to\infty$. If it starts above the stable manifold it eventually lies in 
$G_4$.

\noindent
{\bf Proof} Consider the derivative of the right hand side of the system at 
$S_1$. This matrix has trace $-\alpha-10y-1<0$ and determinant
$\alpha (1-2y)<0$. Thus it has one positive and one negative eigenvalue. Its 
stable manifold $V_s$ is one-dimensional and lies in $G_2\cup G_3$. Along this 
manifold $\frac{\dot y}{\dot x}$ is negative. It follows that $V_s$ is the 
graph of a function of $x$. As $x$ decreases along the part of $V_s$ to the 
left of $S_1$ the derivative of this function remains bounded and so $V_s$ 
intersects the $x$-axis. As $x$ increases along the part of $V_s$ to the right 
of $S_1$ the derivative of the function remains bounded away from zero and so 
$V_s$ intersects the $x$-axis. It can be concluded that the complement of 
$V_s$ in the positive orthant has two connected components $H_1$ and $H_2$, 
where $H_1$ has compact closure. A point is said to lie below the stable 
manifold if it belongs to $H_1$ and above the stable manifold if it belongs to 
$H_2$. A solution which starts in one of these two components remains in it. A
solution which starts in $H_1$ cannot reach $G_4$ and one which starts in
$H_2$ cannot reach $G_1$. Thus Lemma 2 follows from Lemma 1. $\blacksquare$

\noindent
{\bf Lemma 3} A solution of (\ref{hahn1})-(\ref{hahn2}) which is eventually
contained in $G_4$ has, after a suitable translation of $t$, the asymptotics 
$x=\frac{\alpha}{5}e^{\frac{\alpha t}{5}}+\ldots$,
$y=\frac{\sqrt{2}\alpha}{5}e^{\frac{\alpha t}{10}}+\ldots$
for $t\to\infty$. 

\noindent
{\bf Proof} If a solution is eventually contained in $G_4$ then
$r=\sqrt{x^2+y^2}$ must tend to infinity as $t\to\infty$. For $r$ is
an increasing function of $t$ and if it were bounded the solution would have to 
converge to a steady state. However there are no steady states in $G_4$.
It then follows from the defining equations for $G_4$ that both $x$ and $y$
tend to infinity as $t\to\infty$. We now consider the Poincar\'e 
compactification of the system \cite{perko01}. Usually this compactification 
is constructed using two charts, covering neighbourhoods of the $x$- and 
$y$-axes respectively. For a solution which is in $G_4$ for $t\ge t_1$ we have 
seen that $y$ tends to infinity for $t\to\infty$ and hence $\frac{x}{y}$ tends 
to infinity. This means that the solution eventually leaves a neighbourhood of 
the origin in the chart covering a neighbourhood of the $y$-axis and lies in 
the chart covering a neighbourhood of the $x$-axis. Moreover it tends to the 
origin in the latter chart as $t\to\infty$. The chart we are talking about is 
defined by the coordinates $X=1/x$ and $Z=y/x$. Define a new time coordinate 
$\tau$ which satisfies $\frac{d\tau}{dt}=x$. The transformed system is
\begin{eqnarray}
&&\frac{dX}{d\tau}=\alpha X^2-3XZ^2,\\
&&\frac{dZ}{d\tau}=2\alpha X+(\alpha-1)XZ-5Z^2-3Z^3.
\end{eqnarray}
Both eigenvalues of the linearization at the origin are zero and so we must
blow up the origin to get more information. This will be done by means of 
a quasihomogeneous blow-up following \cite{dumortier06}. In the notation
used there the exponents calculated using the Newton polygon are $(2,1)$. There
are two transformations to be done, corresponding to the two coordinates. The
first of these is given by the correspondence $(X,Z)=(u^2,uv)$. We have 
\begin{equation}
\frac{dX}{d\tau}=2u\frac{du}{d\tau}=\alpha u^4-3u^4v^2
\end{equation}
and hence
\begin{equation}
\frac{du}{d\tau}=\frac12(\alpha u^3-3u^3v^2).
\end{equation}
Furthermore
\begin{equation}
\frac{dZ}{d\tau}=u\frac{dv}{d\tau}+v\frac{du}{d\tau}
=2\alpha u^2-5u^2v^2+(\alpha-1)u^3v-3u^3v^3
\end{equation}
and hence 
\begin{equation}
\frac{dv}{d\tau}=2\alpha u-5uv^2-(\alpha-1)u^2v+3u^2v^3
-\frac12(\alpha u^3v-3u^3v^3).
\end{equation}
If we now introduce a new time coordinate $s$ satisfying $\frac{ds}{d\tau}=u$
then the system becomes
\begin{eqnarray}
&&\frac{du}{ds}=\frac12(\alpha u^2-3u^2v^2),\\
&&\frac{dv}{ds}=2\alpha -5v^2+(\alpha-1)uv-3uv^3
-\frac12(\alpha u^2v-3u^2v^3).
\end{eqnarray}
When $v=0$ the derivative of $v$ is positive. Thus no solution can have
an $\omega$-limit point on the $u$-axis. Hence the solution must eventually
be contained in the chart defined by the second transformation, which is 
given by $(X,Z)=(uv^2,v)$. In this case
\begin{equation}
\frac{dZ}{d\tau}=\frac{dv}{d\tau}=2\alpha uv^2-5v^2+(\alpha-1)uv^3-3v^3.
\end{equation}
Furthermore
\begin{equation}
\frac{dX}{d\tau}=v^2\frac{du}{d\tau}+2uv\frac{dv}{d\tau}=\alpha u^2v^4-3uv^4.
\end{equation}
and hence
\begin{equation}
\frac{du}{d\tau}=\alpha u^2v^2-3uv^2-4\alpha u^2v+10uv-2(\alpha-1)u^2v^2+6uv^2.
\end{equation}
If we now introduce a new time coordinate $s$ satisfying $\frac{ds}{d\tau}=v$
then the system becomes
\begin{eqnarray}
&&\frac{du}{ds}=10u+3uv-4\alpha u^2-(\alpha-2)u^2v,\\
&&\frac{dv}{ds}=-5v+2\alpha uv+(\alpha-1)uv^2-3v^2.
\end{eqnarray}
The $u$- and $v$-axes are invariant, the origin is a steady state, which is a
hyperbolic saddle, and there is an additional steady state $(5/2\alpha,0)$.
The latter steady state has one negative and one zero eigenvalue. If we 
transform any positive solution then in the blown-up Poincar\'e compactification
it must tend to that point. To get more details we translate the steady state to
the origin using a coordinate transformation. Let $w=u-\frac{5}{2\alpha}$ be a 
new coordinate. Then the equations become
\begin{eqnarray}
&&\frac{dw}{ds}=-10w-4\alpha w^2+3wv+\frac{15}{2\alpha}v
-(\alpha-2)\left(w+\frac{5}{2\alpha}\right)^2v,\\
&&\frac{dv}{ds}=2\alpha wv
+(\alpha-1)\left(w+\frac{5}{2\alpha}\right)v^2-3v^2.
\end{eqnarray}
The first can be rewritten as
\begin{equation}
\frac{dw}{ds}=-10w+\frac{5(\alpha+10)}{4\alpha^2}v-4\alpha w^2
-\frac{2(\alpha-5)}{\alpha}wv-(\alpha-2)w^2v.
\end{equation}
We now apply centre manifold theory (cf. \cite{perko01}, Section 2.7).
The centre manifold can be written in the form 
$w=\frac{\alpha+10}{8\alpha^2}v+r(v)$ with a remainder term $r$ which is at 
least quadratic. Consider the contributions to the right hand side of the 
evolution equation for $v$ which are quadratic in $v$. We get
\begin{eqnarray}
&&\frac{dv}{ds}=\left[\frac{\alpha+10}{4\alpha}+\frac{5\alpha-5}{2\alpha}-3
\right]v^2+\ldots\nonumber\\
&&=-\frac{1}{4}v^2+\ldots
\end{eqnarray}
After translating $s$ if necessary we get $v=\frac{4}{s}+\ldots$. 
Substituting this into the defining equation for $s$ gives $\tau=\frac18 s^2$
and $v=\sqrt{\frac{2}{\tau}}+\ldots$. It follows that 
$X=\frac{5}{\alpha\tau}+\ldots$ and $Z=\sqrt{\frac{2}{\tau}}+\ldots$. Next
we compute the transformation from $\tau$ to $t$. We have 
$\frac{dt}{d\tau}=\frac{5}{\alpha\tau}+\ldots$ and hence up to a translation
of the time coordinate $t=\frac{5}{\alpha}\log\tau+\ldots$ and 
$\tau=e^{\frac{\alpha t}{5}}+\ldots$. When written in the original variables these 
relations give $x=\frac{\alpha}{5}e^{\frac{\alpha t}{5}}+\ldots$,
$y=\frac{\sqrt{2}\alpha}{5} e^{\frac{\alpha t}{10}}+\ldots$. 
$\blacksquare$

\noindent
{\bf Theorem 1} A positive solution of (\ref{hahn1})-(\ref{hahn2}) with 
$\alpha>0$ and $\beta=0$ belongs to one of the following three classes.

\noindent
(i) It starts below the stable manifold of $S_1$ and $x$ and $y$ converge to 
zero as $t\to\infty$.

\noindent
(ii) It starts on the stable manifold of $S_1$ and converges to $S_1$ as 
$t\to\infty$.

\noindent
(iii) It starts above the stable manifold of $S_1$ and $x$ and $y$ tend to 
infinity as $t\to\infty$, with the asymptotics given in Lemma 3.

\noindent
In particular, every bounded solution converges to a steady state as 
$t\to\infty$.

\noindent
{\bf Proof} This is obtained by combining Lemma 1 - Lemma 3. $\blacksquare$

Note that because after transformation to the Poincar\'e compactification
each solution converges to a steady state there exist no periodic solutions.
In other words the system does not exhibit sustained oscillations. The
eigenvalues of the linearization of the system about $S_0$ are real because
the axes are invariant manifolds. The steady state $S_1$ has been shown to be
hyperbolic with the eigenvalues of the linearization about that point being
real. Thus damped oscillations decaying to one of the steady states are also
ruled out.

\section{The case with photorespiration}\label{photoresp}

In this case we consider the full system where $\alpha$ and $\beta$ are both 
non-zero. 

\noindent
{\bf Lemma 4} Corresponding to positive initial data for 
(\ref{hahn1})-(\ref{hahn2}) with $\alpha>0$ and $\beta>0$ given at $t=t_0$ 
there exists a solution on the interval $(t_0,\infty)$ and it is bounded.

\noindent
{\bf Proof} Taking a suitable linear combination of the equations gives
$\frac{d}{dt}(5x+3y)=-\beta x^2+\alpha x-3y$. If $x\ge\alpha/\beta$ then the 
right hand side is negative. If $x\le\alpha/\beta$ then 
$\alpha x\le \alpha^2/\beta$. Thus if also $y\ge\alpha^2/3\beta$ the right 
hand side is negative. If a solution satisfies 
$5x+3y>\beta^{-1}(5\alpha+\alpha^2)$ at some time then it must be in one of 
the regions where the time derivative of $5x+3y$ is negative. Thus the value 
of $5x+3y$ is bounded by the maximum of its initial value and 
$\beta^{-1}(5\alpha+\alpha^2)$. It follows that all solutions of this system 
can be extended to exist globally in the future and are bounded. $\blacksquare$

Consider now steady states of (\ref{hahn1})-(\ref{hahn2}). 

\noindent
{\bf Lemma 5}  

\noindent
(i) For $\alpha^2/\beta<32$ the only non-negative steady state is
the origin, which we once again denote by $S_0$.

\noindent
(ii) For $\alpha^2/\beta=32$ there is precisely one positive steady state,
which we call $S_1$.

\noindent
(iii) For $\alpha^2/\beta>32$ there are precisely two positive steady 
states. For one of these, which we call $S_1$, both coordinates are smaller 
than the corresponding coordinates of the other steady state, which we call 
$S_2$. 

\noindent
{\bf Proof} Any steady state satisfies $\alpha x=y^2+2y$, so 
that its $y$-coordinate determines its $x$-coordinate. In fact the 
$y$-coordinate is a monotone function of the $x$-coordinate. At the same time 
$\beta x^2=y^2-y$. Squaring the first of these equations and
substituting it into the second gives
\begin{equation}
\beta (y^2+2y)^2=\alpha^2 (y^2-y)
\end{equation}
and hence
\begin{equation}
y[\beta y^3+4\beta y^2+(4\beta-\alpha^2)y+\alpha^2]=yp(y)=0.
\end{equation}
The positive steady states are in one to one correspondence with the positive
roots of the cubic $p(y)$. Since
$p(0)>0$ there is at least one negative root and there are at most two positive 
roots. If $4\beta-\alpha^2>0$ then it follows from Descartes' rule of signs
\cite{murray89} that there are no positive roots. Further information about 
the number of positive roots can be obtained by looking at the discriminant of 
the polynomial $p$. It is given by 
\begin{equation}
\Delta=\alpha^2\beta[96\beta^2-131\alpha^2\beta+4\alpha^4]. 
\end{equation}
There are two positive values of $\alpha^2/\beta$ for which $\Delta$ vanishes,
namely $\zeta_{\pm}=\frac{131\pm 125}{8}$. We have $\zeta_-=\frac34$ and 
$\zeta_+=32$. When $\Delta<0$ the polynomial $p$ has only one real root and it 
must be negative. When $\Delta\ge 0$ all roots are real. Only in this case can 
there be more than one positive root. For $\alpha^2/\beta=32$ there is a root 
which is at least double. If this root were negative then it would follow that 
$4\beta-\alpha^2>0$, a contradiction. Thus the double root is positive. For 
$\alpha^2/\beta>32$ there are two positive roots. $\blacksquare$

\begin{figure}[ht]
\begin{center}
\includegraphics[scale=0.4]{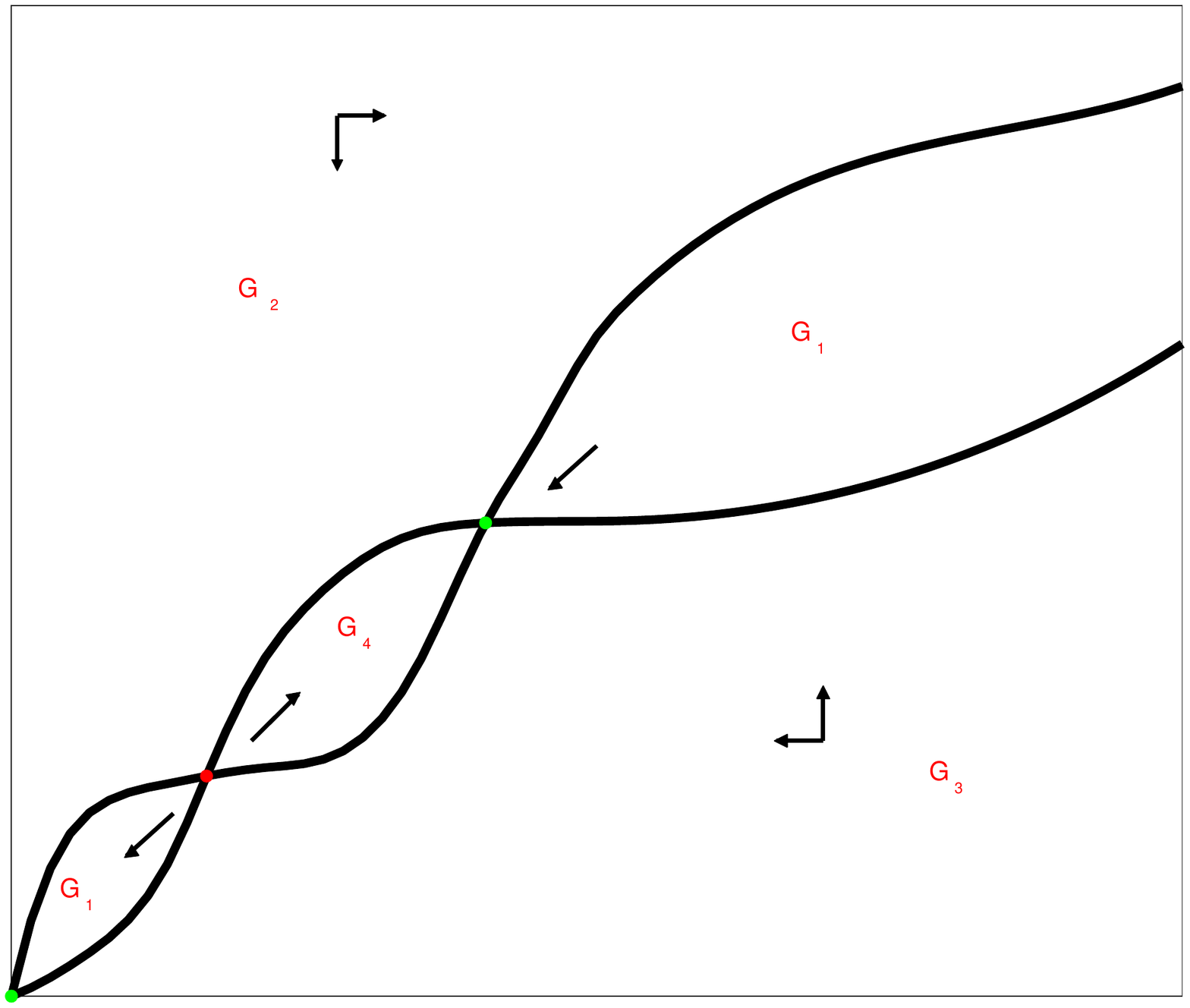}
\end{center}
\caption{Phase portrait in the presence of photorespiration}
\label{fig:photoresp}
\end{figure}

Both of the nullclines of this system are of the form $f(x)=g(y)$ for 
monotone increasing functions $f$ and $g$. Thus we can write them as graphs of 
functions of $x$ or of functions of $y$. Due to Lemma 5 we know that the two 
nullclines intersect at the origin and at no, one or two points in the 
positive region, depending on the parameters. It can then be concluded that 
when there are no, one and two positive steady states the complement of the 
union of the nullclines has three, four and five connected components, 
respectively. As in the previous section these components can be labelled with 
the signs of $\dot x$ and $\dot y$. In case (i) of Lemma 5 there is one 
component with each of the labels $(-,-)$, $(-,+)$ and $(+,-)$. In case (ii)
there are two components with the label $(-,-)$ and one component with each of 
the labels $(-,+)$ and $(+,-)$. In case (iii) (cf. Figure~\ref{fig:photoresp})
there are two components with
the label $(-,-)$ and one component with each of the labels $(-,+)$, $(+,-)$
$(+,+)$. The components of the complements of the set of steady states in
the nullclines can be labelled as $(-,0)$, $(0,-)$, $(+,0)$ and $(0,+)$.

At a point where the nullclines cross they are tangent if and only if the 
linearization at the point has a zero eigenvalue. This happens precisely when 
the determinant of the linearization at that point is zero. This is the only
way that a steady state can fail to be hyperbolic since the fact that the trace
is negative rules out the possibility of a pair of complex conjugate imaginary
eigenvalues. With the information we already have about steady states it can 
be concluded that the determinant is zero precisely when
\begin{equation}
q(y)=4\beta y^3+12\beta y^2 +(8\beta-2\alpha^2)y+\alpha^2=0.
\end{equation}
Taking suitable linear combinations of the equations $p(y)=0$ and $q(y)=0$ 
leads to two quadratic equations for $y$. These can be combined to give a
linear equation for $y$ and substituting this back into $r(y)$ leads to
the equation 
\begin{equation}
9\gamma (-4\gamma^2+131\gamma-96)=0
\end{equation}
where $\gamma=\frac{\alpha^2}{\beta}$. It follows that all steady states are 
hyperbolic when $\gamma>32$. It can be concluded
that except in case (ii) the steady states are hyperbolic. In case (ii) the 
linearization at $S_1$ has one zero and one negative eigenvalue. In case (iii)
it has one positive and one negative eigenvalue. 

\noindent
{\bf Theorem 2} Any positive solution of (\ref{hahn1})-(\ref{hahn2}) with 
$\alpha>0$ and $\beta>0$ converges a steady state as $t\to\infty$. If 
$\frac{\alpha^2}{\beta}<32$ there are no points $S_1$ and $S_2$ and all 
solutions converge to $S_0$. If $\frac{\alpha^2}{\beta}=32$ there is no 
point $S_2$ and points above or on and below the unique centre manifold of 
$S_1$ converge to $S_1$ and $S_0$ respectively. If $\frac{\alpha^2}{\beta}>32$ 
then points above, on or below the stable manifold of $S_1$ converge to $S_2$, 
$S_1$ and $S_0$ respectively. 

\noindent
{\bf Proof} Note that the stable manifold of $S_1$ is always one-dimensional. 
By the same arguments as in the case $\beta=0$ it can be shown that this 
manifold is the graph of a function of $x$ and that its complement is the 
union of two components $H_1$ and $H_2$. Components with the sign combination 
$(-,-)$ have boundaries with the sign combinations $(-,0)$ and $(0,-)$. Using 
the information on the signs of $\ddot x$ and $\ddot y$ shows that these 
components are invariant. For instance, a solution which satisfies $\dot x=0$ 
and $\dot y<0$ at some point satisfies $\ddot x<0$. Similarly components with 
the sign combination $(+,+)$ are invariant.
It can be shown as in the case $\beta=0$ that 
any solution which starts in a component with one of sign combinations $(+,-)$ 
or $(-,+)$ and does not converge to a steady state as $t\to\infty$ must enter 
one of the components with the sign combination $(-,-)$ or $(+,+)$ after a 
finite time. Once it enters a component of this type it must stay there and 
converge to a steady state as $t\to\infty$. Thus every solution converges to a 
steady state as $t\to\infty$. It is then straightforward to determine which 
steady state it converges to in different cases. $\blacksquare$

Note that since every solution converges to a steady state the system exhibits
no sustained oscillations. With the information we have about the linearization
about the fixed points we can argue as in the previous section that there are
no damped oscillations close to the point $S_0$. Using Lemma A2 of the Appendix
we get the corresponding conclusion for $S_1$ and $S_2$.

Consider what happens if $\beta$ tends to zero while $\alpha$ has a fixed
positive value. The polynomial converges. In the limit there is a unique
root, which is $S_1$. It is a hyperbolic saddle. Thus it is the limit of
a steady state of the system in the general case as $\beta\to 0$. The 
approximating solution must coincide with the point $S_1$ in the general
system. Let $\gamma=\frac{\alpha^2}{\beta}$ and define $z=\gamma^{-1/2}y$
and $q(z)=p(y)$. Then if $\beta$ tends to zero while $\alpha$ has a fixed
positive value the polynomial $q$ converges. Again there is a unique 
positive root in the limit with $z=1$. It is approximated by a root for positive
$\beta>0$ and that corresponds to the steady state $S_2$. We conclude that
as $\beta$ tends to zero the coordinates of $S_2$ have the asymptotic behaviour
$x=\alpha\beta^{-1}+\ldots$ and $y=\alpha\beta^{-\frac12}+\ldots$.

\section{The system with the fifth power}\label{fifth}

In this section we study the system where $y^2$ is replaced in 
(\ref{hahn1})-(\ref{hahn2}) by $y^5$. This is 
\begin{eqnarray}
&&\frac{dx}{dt}=-\alpha x-2\beta x^2+3y^5,\label{hahn51}\\
&&\frac{dy}{dt}=2\alpha x+3\beta x^2-5y^5-y.\label{hahn52}
\end{eqnarray}
The aim is to see to what extent the results obtained for 
(\ref{hahn1})-(\ref{hahn2}) generalize to (\ref{hahn51})-(\ref{hahn52}).

In the case $\alpha=0$ the analysis of (\ref{hahn1})-(\ref{hahn2}) extends
without essential changes to (\ref{hahn51})-(\ref{hahn52}) to give the same 
qualitative results. When $\alpha\ne 0$ and $\beta=0$ the analysis up to and
including Lemma 2 extends easily. Of course the explicit formulae in the 
definitions of the invariant regions $G_i$ are modified by replacing $y^2$ by
$y^5$. 

\noindent
{\bf Lemma 6} A solution of (\ref{hahn51})-(\ref{hahn52}) which is eventually
contained in $G_4$ has, after a suitable translation of $t$, the asymptotics 
$x=(\frac{4\alpha}{5})^{\frac14}e^{\frac{\alpha t}{5}}+\ldots$,
$y=2^{\frac{1}{20}}\left(\frac{2\alpha}{5}\right)^{\frac{1}{4}}e^{\frac{\alpha t}{25}}
+\ldots$ for $t\to\infty$.

\noindent
{\bf Proof} That the arguments from the case of (\ref{hahn1})-(\ref{hahn2})
extend easily is also true of the first part of the proof of Lemma 3 which 
shows that the late time behaviour can be analysed in one of the charts of the 
Poincar\'e compactification. In this case the time coordinate must be rescaled 
in a different way from what was done previously. Let $\frac{d\tau}{dt}=x^4$. 
In the case of 
(\ref{hahn51})-(\ref{hahn52}) the transformation to this chart gives
\begin{eqnarray}
&&\frac{dX}{d\tau}=\alpha X^5-3XZ^5,\\
&&\frac{dZ}{d\tau}=2\alpha X^4+(\alpha-1)X^4Z-5Z^5-3Z^6.
\end{eqnarray} 
The linearization of the system at the origin is identically zero. To get
more information we do a quasi-homogeneous blow-up. The exponents 
calculated using the Newton polygon are $(5,4)$. Once again, there are two
transformations to be done. The first of these is given by the correspondence 
$(X,Z)=(u^5,u^4v)$. We have 
\begin{equation}
\frac{dX}{d\tau}=5u^4\frac{du}{d\tau}=\alpha u^{25}-3u^{25}v^5
\end{equation}
and hence
\begin{equation}
\frac{du}{d\tau}=\frac15(\alpha u^{21}-3u^{21}v^5).
\end{equation}
Furthermore
\begin{equation}
\frac{dZ}{d\tau}=u^4\frac{dv}{d\tau}+4u^3v\frac{du}{d\tau}
=2\alpha u^{20}-5u^{20}v^5+(\alpha-1)u^{24}v-3u^{24}v^6
\end{equation}
and hence 
\begin{equation}
\frac{dv}{d\tau}=2\alpha u^{16}-5u^{16}v^5-(\alpha-1)u^{20}v+3u^{20}v^6
-\frac45(\alpha u^{20}v-3u^{20}v^6).
\end{equation}
If we now introduce a new time coordinate $s$ satisfying 
$\frac{ds}{d\tau}=u^{16}$ then the system becomes
\begin{eqnarray}
&&\frac{du}{ds}=\frac15(\alpha u^5-3u^5v^5),\\
&&\frac{dv}{ds}=2\alpha -5v^2+(\alpha-1)u^4v-3u^4v^6
-\frac45(\alpha u^4v-3u^4v^6).
\end{eqnarray}
Just as in the case with the quadratic nonlinearity we see that the solution
must eventually be contained in the chart defined by the second transformation
which is given by $(X,Z)=(uv^5,v^4)$. In that case
\begin{equation}
\frac{dZ}{d\tau}=4v^3\frac{dv}{d\tau}=2\alpha u^4v^{20}-5v^{20}
+(\alpha-1)u^4v^{24}-3v^{24}.
\end{equation}
Hence
\begin{equation}
\frac{dv}{d\tau}=\frac14\left\{2\alpha u^4v^{17}-5v^{17}+(\alpha-1)u^4v^{21}
-3v^{21}\right\}.
\end{equation}
Furthermore
\begin{equation}
\frac{dX}{d\tau}=v^5\frac{du}{d\tau}+5uv^4\frac{dv}{d\tau}
=\alpha u^5v^{25}-3uv^{25}.
\end{equation}
and hence
\begin{equation}
\frac{du}{d\tau}=\alpha u^5v^{20}-3uv^{20}
-\frac54\{2\alpha u^5v^{16}-5uv^{16}+(\alpha-1)u^5v^{20}-3uv^{20}\}.
\end{equation}
In terms of the time coordinate $s$ with $\frac{ds}{d\tau}=v^{16}$ we get
\begin{eqnarray}
&&\frac{du}{ds}=\alpha u^5v^4-3uv^4
-\frac54\{2\alpha u^5-5u+(\alpha-1)u^5v^4-3uv^4\},\\
&&\frac{dv}{ds}=\frac14\left\{2\alpha u^4v-5v+(\alpha-1)u^4v^5
-3v^5\right\}.
\end{eqnarray}
The axes are invariant and the origin is a hyperbolic saddle. There is a steady
state at the point $(u_0,0)$ with $u_0=\left(\frac{5}{2\alpha}\right)^{\frac14}$.
If we transform any solution then in the blown-up Poincar\'e compactification it
must converge to this point. To get more details we translate the steady state
to the origin using a coordinate transformation. Let $w=u-u_0$. Then the 
equations become
\begin{eqnarray}
&&\frac{dw}{ds}=-\frac12 u_0v^4
-\frac54\{[2\alpha (u_0+w)^4-5](u_0+w)
-\frac{\alpha +5}{2\alpha}u_0v^4\}\nonumber\\
&&+O(v^4w),\label{wevol}\\
&&\frac{dv}{ds}=\frac14\left\{[2\alpha (u_0+w)^4-5]
-\frac{\alpha +5}{2\alpha}v^4\right\}v+O(v^5w).\label{vevol}
\end{eqnarray}
Here we have explicitly retained only those terms which are required for the 
calculation which will now be done. It follows from the definition of the 
centre manifold that $w=h(v)$ for a function $h$ with $h(v)=O(v^2)$. The 
derivative of this relation with respect to time also holds. Hence 
$\dot w=h'(v)\dot v$. It follows from (\ref{vevol}) that $\dot v=O(v^3)$ and 
so $\dot w=O(v^4)$. It follows from (\ref{wevol}) that $w=O(v^4)$. 
Hence $\dot v=O(v^5)$ and $\dot w=O(v^6)$. It can be concluded from the 
evolution equation for $w$ that
\begin{equation}
[2\alpha (u_0+w)^4-5]-\frac{\alpha +5}{2\alpha}v^4=-\frac25 v^4+\ldots.
\end{equation}
It follows that $\frac{dv}{ds}=-\frac{1}{10}v^5+\ldots$. We see that the flow 
on the centre manifold is towards the steady state. After translating $s$ if
necessary we get $v=\left(\frac{5}{2s}\right)^{\frac14}+\ldots$. Substituting 
this into the defining equation for $s$ gives 
$s=5^{\frac15}\left(\frac{5}{2}\right)^{\frac45}\tau^{\frac15}+\ldots$
and $v=\left(\frac{1}{2\tau}\right)^{\frac{1}{20}}+\ldots$. Substituting for
the original variables gives $X=u_0\left(\frac{1}{2\tau}\right)^{\frac14}+\ldots$
and $Z=\left(\frac{1}{2\tau}\right)^{\frac15}+\ldots$. Next we compute the 
transformation from $\tau$ to $t$. We have 
$\frac{dt}{d\tau}=\left(\frac{u_0^{4}}{2\tau}\right)+\ldots$. Hence
$t=\left(\frac{u_0^{4}}{2}\right)\log\tau+\ldots$ and 
$\tau=e^{4\alpha t/5}+\ldots$. Finally we get
$x=(\frac{4\alpha}{5})^{\frac14}e^{\frac{\alpha t}{5}}$ and 
$y=2^{\frac{1}{20}}\left(\frac{2\alpha}{5}\right)^{\frac{1}{4}}e^{\frac{\alpha t}{25}}$.
$\blacksquare$

\noindent
{\bf Theorem 3} A positive solution of (\ref{hahn51})-(\ref{hahn52}) with 
$\alpha>0$ and $\beta=0$ belongs to one of the following three classes.

\noindent
(i) It starts below the stable manifold of $S_1$ and $x$ and $y$ converge to 
zero as $t\to\infty$.

\noindent
(ii) It starts on the stable manifold of $S_1$ and converges to $S_1$ as 
$t\to\infty$.

\noindent
(iii) It starts above the stable manifold of $S_1$ and $x$ and $y$ tend to 
infinity as $t\to\infty$, with the asymptotics given in Lemma 6.

\noindent
In particular, every bounded solution converges to a steady state as 
$t\to\infty$.

\noindent
{\bf Proof} The proof is identical to that of Theorem 1 except that Lemma 3
is replaced by Lemma 6. $\blacksquare$

It is interesting to compare the asymptotics in Lemma 6 with those obtained
in \cite{rendall14} for a more elaborate model of the Calvin cycle. In 
Lemma 6 we see that both unknowns have growing exponential asymptotics but 
that the exponent for GAP is one fifth of that for the other variables. The main
system considered in \cite{rendall14} has five unknowns and has solutions for
which all unknowns have growing exponential asymptotics. In that case the 
exponent for GAP is one fifth of that for the other four unknowns. These 
four unknowns satisfy a system of the form $\frac{d\bar x}{dt}=A\bar x+R$ where 
$R$ is considered as a remainder term and the larger exponent is an eigenvalue 
of $A$. There is a natural analogue of this equation for the system 
(\ref{hahn51})-(\ref{hahn52}) with $\beta=0$. It is the equation
$\frac{d}{dt}(5x+3y)=\frac{\alpha}{5}(5x+y)-\left(\frac{\alpha}{5}+1\right)y.$ 
Here the last term is to be considered as the remainder. Note that in the 
asymptotics of Lemma 6 $y$ is much smaller than $x$ at late times so that this  
treatment as remainder term is reasonable. Since there is ony one unknown
growing at the maximal rate in this case the matrix $A$ is replaced by a number
and that number is $\alpha/5$. Thus we see that on a heuristic level the 
exponents in the two cases agree. The statement of Theorem 3 is stronger than
the analogous statement in \cite{rendall14} in the following sense. The
description of the asymptotic behaviour in \cite{rendall14} is only obtained
for some non-empty subset of initial data which is not further characterized
while the set of initial data giving rise to this asymptotic behaviour in
Theorem 3 is much more explicit. 

Consider next the case where the coefficients $\alpha$ and $\beta$ in
(\ref{hahn51})-(\ref{hahn52}) are both positive. The solutions are bounded
using the same argument as in the proof of Lemma 4. As in the case of 
(\ref{hahn1})-(\ref{hahn2}) the nullclines are of the form $f(x)=g(y)$ for 
monotone increasing functions $f$ and $g$. A steady state satisfies the
equations
\begin{eqnarray}
&&y=\frac13 (\alpha x-\beta x^2),\\
&&x=\frac{1}{\alpha}(y^5+2y).
\end{eqnarray}
Substituting the second of these equations into the first gives
\begin{equation}
p(y)=\beta y^{10}+4\beta y^6-\alpha^2 y^5+4\beta y^2+\alpha^2 y=0.
\end{equation}
By Descartes' rule of signs this equation can have at most two positive 
solutions. Since the derivative of the polynomial $p$ at zero is positive 
$p(y)>0$ for $y$ slightly larger than zero. Thus if $p(y)<0$ for some $y>0$
the polynomial has two positive roots. Now $p(2)=1296\beta-28\alpha^2$.
Thus for fixed $\beta$ if $\alpha$ is large enough we have $p(2)<0$ and 
$p$ has two positive roots. If we define values of $x$ corresponding to these
two values of $y$ we obtain two positive steady states of the system
(\ref{hahn51})-(\ref{hahn52}). On the other hand, if $\beta>\alpha^2$ there
are no positive steady states. 

We have not succeeded in obtaining information about the hyperbolicity of 
steady states of this system which is as complete as that which we obtained in 
the case of a quadratic nonlinearity. It is, however, possible to show that 
for generic values of the parameter $\gamma$ all steady states are hyperbolic. 
We can calculate polynomials $p$ and $q$ as in the case with a quadratic 
nonlinearity but it is not possible to solve explicitly for their common
roots $y$. Instead we 
can proceed as follows. For any non-hyperbolic steady state we obtain 
equations of the form
\begin{eqnarray}
&&p(y)=p_1(y)-\gamma (y-1)=0,\nonumber\\
&&q(y)=q_1(y)-\gamma(5\gamma^4-1)=0
\end{eqnarray}
for certain polynomials $p_1$ and $q_1$ which do not depend on $\gamma$. Hence
\begin{equation}
s(y)=(5\gamma^4-1)p_1(y)-(y-1)q_1(y)=0.
\end{equation}
Since the polynomial $s$ is non-constant this equation has only finitely many 
solutions $y$. For any given solution $y$ there is at most one corresponding
value of $\gamma$. Hence for all but finitely many values of $\gamma$ all
steady states are hyperbolic. 

With the information on steady states just obtained we can prove an analogue of
Theorem 2 for the system with the fifth power using the same techniques. The
result is

\noindent
{\bf Theorem 4} Any positive solution of (\ref{hahn51})-(\ref{hahn52}) with 
$\alpha>0$ and $\beta>0$ converges to a steady state as $t\to\infty$. If 
$\frac{\alpha^2}{\beta}<1$ there are no points $S_1$ and $S_2$ and all 
solutions converge to $S_0$. If $\frac{\alpha^2}{\beta}$ is large enough 
then points above, on or below the stable manifold of $S_1$ converge to $S_2$, 
$S_1$ and $S_0$ respectively.

By scaling the unknowns $x$ and $y$ by the same factor and $t$ by another 
factor it is possible to transform the more general system 
\begin{eqnarray}
&&\frac{dx}{dt}=-\alpha x-2\beta x^2+3Ay^5,\label{hahn51r}\\
&&\frac{dy}{dt}=2\alpha x+3\beta x^2-5Ay^5-By.\label{hahn52r}
\end{eqnarray} 
for general positive constants $A$ and $B$ into the system 
(\ref{hahn51})-(\ref{hahn52}). Thus the results obtained for 
(\ref{hahn51})-(\ref{hahn52}) imply analogous results for 
(\ref{hahn51r})-(\ref{hahn52r}). This observation will be used in the next 
section.

\section{Derivation from the three-dimensional system}\label{3d}

The system (\ref{hahn1})-(\ref{hahn2}) was derived by Hahn from a 
three-dimensional system but he did not give a mathematical formulation of
the relation between the two systems. The three-dimensional system is, in
a modified notation,
\begin{eqnarray}
&&\frac{dx}{dt}=-k_1 x-2k_2 x^2+3k_4 z^5,\\
&&\frac{dy}{dt}=2k_1 x+3k_2 x^2-k_3 y,\\
&&\frac{dz}{dt}=k_3 y-5k_4 z^5-(k_5+k_6)z.
\end{eqnarray}
We now consider a limit where $k_3$ becomes large. This means that the reaction
producing triose phosphate from PGA is very fast. Let 
$k_3=\epsilon^{-1}\tilde k_3$ and introduce a new variable by $w=y+z$. Then
the equations above are equivalent to the system
\begin{eqnarray}
&&\frac{dx}{dt}=-k_1 x-2k_2 x^2+3k_4 z^5,\\
&&\frac{dw}{dt}=2k_1 x+3k_2 x^2-5k_4 z^5-(k_5+k_6)z,\\
&&\epsilon\frac{dz}{dt}=\tilde k_3 (w-z)-5\epsilon k_4 z^5
-\epsilon (k_5+k_6)z.
\end{eqnarray}
This is a fast-slow system in standard form with fast variable $z$ and slow
variables $x$ and $w$. The critical manifold is given by $z=w$ and the
slow system is 
\begin{eqnarray}
&&\frac{dx}{dt}=-k_1 x-2k_2 x^2+3k_4 w^5,\\
&&\frac{dw}{dt}=2k_1 x+3k_2 x^2-5k_4 w^5-(k_5+k_6)w.
\end{eqnarray}
Replacing $w$ by $y$ and setting $k_1=\alpha$, $k_2=\beta$, $k_4=A$ and 
$k_5+k_6=B$ gives the system (\ref{hahn51r})-(\ref{hahn52r}). The critical
manifold is normally hyperbolic and the one normal eigenvalue is negative.

We know that for certain values of the parameters the system 
(\ref{hahn51})-(\ref{hahn52}) has three steady states $S_0$, $S_1$ and $S_2$.
Moreover, for generic values of $\gamma$ the steady states $S_0$ and $S_2$ are 
hyperbolic sinks while $S_1$ is a hyperbolic saddle with a one-dimensional 
stable manifold. There are heteroclinic orbits connecting $S_0$ to $S_1$ and 
$S_1$ to $S_2$. Putting this
together with the fact that the normal eigenvalue is negative shows that
for suitable parameters with $\epsilon$ small the three-dimensional system
has three steady states $S_0$, $S_1$ and $S_2$ which converge to those with the
corresponding names as $\epsilon=0$. Moreover $S_0$ and $S_2$ are hyperbolic 
sinks and $S_1$ is a hyperbolic saddle with a two-dimensional unstable 
manifold.  

\section{Conclusions and outlook}\label{outlook}

In this paper we have obtained detailed information on minimal models of the
Calvin cycle introduced by Hahn in \cite{hahn91}. A rather complete analysis
of the two-dimensional models of Hahn was given. The relation of the 
two-dimensional to the three-dimensional model of Hahn was discussed but a
comprehensive analysis of the three-dimensional model, which is likely to be
complicated, was postponed to future work. The models in \cite{hahn91}
originated by formal simplification of earlier models due to the same author.
The first is a model with 19 chemical species defined in \cite{hahn84}. It
did not implement a detailed description of photorespiration and a description
of this kind was added in the model of \cite{hahn87}, with 33 species. In the
future it would be desirable to put the understanding of the relations between
these different models on a better mathematical footing. This should also 
allow conclusions about the dynamics of the 
higher-dimensional systems to be obtained. Note that there are some general
references in the literature about the inheritance of dynamical features from
reduced systems (see e.g. \cite{banaji18}, \cite{feliu13}). 

Another interesting task is to relate the models of Hahn to other models of 
the Calvin cycle in the literature. Possible relations to a model of 
Grimbs et al. \cite{grimbs11} studied in \cite{rendall14} were already 
mentioned is section \ref{fifth} and perhaps these could be extended so as to 
give a wider view of runaway solutions of models for the Calvin cycle, i.e 
those solutions where all concentrations tend to infinity. One task is to 
obtain some kind of characterization of models admitting solutions of this 
type. Another is to obtain formulae for the asymptotics of these solutions 
in the case that they do occur. This kind of behaviour can be ruled out if the 
model admits a suitable conservation law. This is, for instance, the case in a 
model of \cite{petterson88} whose mathematical properties were studied in 
\cite{moehring17}. In the model of Hahn with photorespiration boundedness of 
solutions is obtained without there being a conservation law. Another example 
of this is given by a model studied in Section 6 of \cite{rendall14} where the 
original model of Grimbs at al. is modified by including the concentration of 
ATP explicitly. 

It also remains to obtain a comprehensive understanding of solutions where 
some concentrations tend to zero at late times. As discussed in 
\cite{moehring17} this can be related to the biological phenomenon of overload 
collapse. This means intuitively that the production of sugar by the cycle 
cannot meet the demand for export from the chloroplast. In \cite{moehring17}
it was shown that there are solutions of the model of \cite{petterson88}
which admit this phenomenon while in a modification of the model due to 
Poolman \cite{poolman99} these solutions are eliminated. The model of
\cite{poolman99} does not include photorespiration but does include the 
mobilization of glucose from starch. 

In this paper various aspects of the dynamics of the minimal model of Hahn for
the Calvin cycle have been analysed. The reduction to a model with only two 
variables makes it possible to get a good overview of the dynamics. We
believe that this establishes a good starting point for understanding the 
dynamics of more detailed models of the Calvin cycle in the future and we
have indicated some directions in which this could progress. 

\section{Appendix: nullcline analysis}

In this section we discuss how nullclines can be used to obtain information
about the global behaviour of solutions of a two-dimensional dynamical system.
Consider the following system of ordinary differential equations.
\begin{eqnarray}
&&\dot x=f(x,y),\\
&&\dot y=g(x,y)
\end{eqnarray}
where the functions $f$ and $g$ are $C^1$ and defined on an open subset 
$U\subset\R^2$. The nullclines $N_1$ and $N_2$ are the zero sets of $f$ and 
$g$ respectively. Let $G=U\setminus (N_1\cup N_2)$. The open set $G$ is a 
countable union of connected components $G_i$. In what follows we will restrict
to the case that the following assumption is satisfied.

\noindent
{\bf Assumption 1} The complement of the nullclines has only finitely many
connected components. 
 
\noindent
When the system satisfies Assumption 1 it defines a directed graph as follows. 
There is one node for each component $G_i$ and there is a directed edge from 
the node corresponding to $G_i$ to that corresponding to $G_j$ when there is a 
solution which starts from a point of $G_i$ and later enters $G_j$ without 
entering any component of $G$ other than $G_i$ and $G_j$ at an intermediate 
time. Let us call this the succession graph. A cycle in a 
directed graph is a finite sequence of directed edges such that the initial 
node of each edge is the final node of the previous one and the final node of 
the last edge is the initial node of the first one. We now restrict to the 
case that the following assumption is satisfied.

\noindent
{\bf Assumption 2} There exist only finitely many steady states. Whenever a
steady state $S_i$ is in the closure of a component $G_j$ there is a 
continuous curve joining a point of $G_j$ to $S_i$ which does not intersect
any other $G_k$.


\noindent
{\bf Lemma A1} Consider a solution $(x(t),y(t))$ on a time interval $[t_0,t_1)$
which lies in $G_i$ for some $i$ when $t=t_0$ and which lies entirely in 
$\bar G_i$. Then $x(t)$ and $y(t)$ are monotone. They are strictly monotone
as long as the solution lies in $G_i$. Suppose that $t_1$ is 
maximal. If the solution is bounded then it converges to a point $(x^*,y^*)$ 
for $t\to t_1$ which is either a point of $N_1\cup N_2$ or a point of 
$\bar G\setminus G$. If $(x^*,y^*)\in G$ then $(x^*,y^*)\in N_1\cap N_2$ if and 
only if $t_1=\infty$.

\noindent
{\bf Proof} On a component $G_i$ the signs of $\dot x$ and $\dot y$ are 
constant and this implies the monotonicity statements. It follows that 
the limits of $x(t)$ and $y(t)$ as $t\to\infty$ exist, either as real numbers
or as $\pm\infty$. If the solution is bounded then these limits are real
numbers $x^*$ and $y^*$. The point $(x^*,y^*)$ belongs to the closure of $G$. 
Suppose now that $t_1$ is maximal and that $(x^*,y^*)\in G$. We claim that if
$t_1=\infty$ then $(x^*,y^*)$ is a point of $N_1\cap N_2$, and hence a steady
state. Otherwise at least one of $\dot x$ or $\dot y$ would tend to a non-zero 
value, say $c$, as $t\to t_1$. Suppose w.l.o.g. that $\dot x$ has this 
property and that $c>0$. It follows that if $t_2$ is sufficiently 
large then $\dot x(t)\ge \frac12 c t$ for all $t\ge t_2$. Thus $x$ is unbounded,
a contradiction. We conclude that if the interval $[t_0,t_1)$ is infinite 
$(x^*,y^*)\in N_1\cap N_2$. Suppose now conversely that 
$(x^*,y^*)\in N_1\cap N_2$. If $t_1$ were finite it would be possible to extend 
the solution beyond $t=t_1$. But then this solution would have to coincide 
with the time-independent solution $x(t)=x^*$, $y(t)=y^*$ a contradiction. 
Thus if $(x^*,y^*)\in N_1\cap N_2$ the interval $[t_0,t_1)$ is infinite.

\noindent
{\bf Lemma A2} Let $S_i$ be a steady state. Suppose that Assumption 2 is 
satisfied, that there is more than one component $G_j$ having $S_i$ as a 
limit point and that there is no cycle in the succession graph. Then there 
is no damped oscillation converging to $S_i$.
 
\noindent
{\bf Proof} Suppose there is a solution exhibiting a damped oscillation. Due
to Assumption 2 it must intersect each $G_j$ having $S_i$ as a limit point 
more than once. Hence the succession graph contains a cycle, a contradiction.

\end{document}